\documentclass[conference,letterpaper]{IEEEtran}
\usepackage[letterpaper, left=54pt, right=54pt, bottom=54pt, top = 54pt]{geometry}
\newgeometry{left=54pt, right=54pt, bottom=54pt, top = 72pt}
\IEEEoverridecommandlockouts
\usepackage{amsmath}
\usepackage{amsfonts}
\usepackage{hyperref}
\usepackage{subcaption}

\usepackage{tikz}
\usepackage{xfrac}
\usepackage{siunitx}
\usepackage{float}
\usepackage{pgfplots}
\usepackage{tikz-3dplot}
\usetikzlibrary{calc}
\pgfplotsset{width=7cm,compat=1.8}

\usepackage{color}
\usepackage{soul}

\usepackage[style=ieee, 
citestyle=numeric-comp,
sorting=none]{biblatex}
\def\BibTeX{{\rm B\kern-.05em{\sc i\kern-.025em b}\kern-.08em
    T\kern-.1667em\lower.7ex\hbox{E}\kern-.125emX}}
\begin{document}

\title{Models and Predictive Control for Nonplanar Vehicle Navigation\\
\thanks{Videos can be found at \href{https://youtu.be/KHP0uUW4oHY}{https://youtu.be/KHP0uUW4oHY}}
\thanks{Source code: \href{https://github.com/thomasfork/Nonplanar-Vehicle-Control}{https://github.com/thomasfork/Nonplanar-Vehicle-Control}}
\thanks{\textsuperscript{1}Thomas Fork and Francesco Borrelli are with the Department of Mechanical Engineering, University of California, Berkeley, USA \texttt{\{fork, fborrelli\}@berkeley.edu}}
\thanks{\textsuperscript{2}H. Eric Tseng is with the Ford Motor Company}
}

\author{\IEEEauthorblockN{Thomas Fork\textsuperscript{1} }
\and
\IEEEauthorblockN{H. Eric Tseng\textsuperscript{2} }
\and
\IEEEauthorblockN{Francesco Borrelli\textsuperscript{1} }
}

\maketitle

\begin{abstract}
We present a simplified model of a vehicle driving on a nonplanar road. A parametric surface is used to describe the nonplanar road which can describe any combination of curvature, bank and slope. We show that the proposed modeling approach generalizes planar vehicle models that reference a centerline, such as the Frenet model.

We use the proposed approach for vehicle path planning and following using model predictive control. We also model and control vehicle contact with the road surface. 
We demonstrate that the proposed controller improves speed and lane following on complex roads compared to planar vehicle controllers, and mitigates loss of control on complex road surfaces including off-camber turns.
\end{abstract}

%%%%%%%%%%%%%%%%%%%%%%%%%%%%%%%%%%%%%%%%%%%%%%%%%%%%%
\section{Introduction}
Vehicle motion planning models are widely used for control and planning of autonomous vehicles~\cite{7490340} however most control oriented models are planar, i.e. the road surface is confined to a single plane~\cite{7225830}. These are sufficient to describe motion along curved flat roads~\cite{MICAELLI1994249}, banked roads~\cite{bank_angle_paper} and sloped roads~\cite[Ch.~4]{rajamani_book}, but not a combination of the three or motion along a path with varying bank and slope. The objective of this paper is to derive a simplified three dimensional vehicle model for the nonplanar case and use it to design model predictive control (MPC) algorithms. 

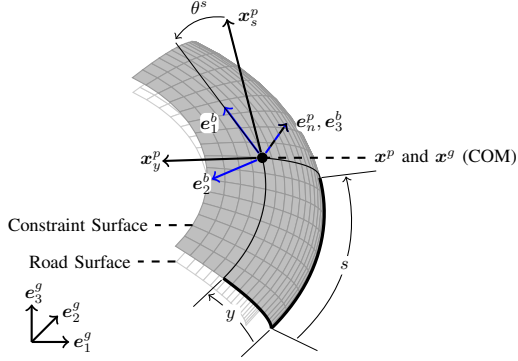
\begin{figure}[t]%{0.45\textwidth}
    \centering
    \begin{tikzpicture}[inner sep=0pt]

    \node[anchor = west] (g1) at (.55,0){\scalebox{0.7}{$\boldsymbol{e}_1^g$}};
    \node[anchor = west] (g2) at (.4,.4){\scalebox{0.7}{$\boldsymbol{e}_2^g$}};
    \node[anchor = west] (g3) at (-0.1,.65){\scalebox{0.7}{$\boldsymbol{e}_3^g$}};
    \draw[thick,->](0,0)--(0,.5);
    \draw[thick,->](0,0)--(0.35,.35);
    \draw[thick,->](0,0)--(.5,0);
    
    \begin{axis}[axis line style={draw=none},
          view={35}{75},
          tick style={draw=none},
          xticklabels={,,},
          yticklabels={,,},
          zticklabels={,,},
          clip=false
            ]
    \addplot3[
      surf,
      colormap = {graymap}{color = (white) color = (white)},
      shader=faceted,
      samples = 17,
      domain=-0.66:0,
      domain y= 0 : pi/2,
      opacity = 1
    ]
    ({sqrt(.8-x^2) * cos(deg(y))},
     {sqrt( .8-x^2 ) * sin(deg(y))},
     {-x-0.2});
     
    \addplot3[
      surf,
      colormap = {graymap}{color = (lightgray) color = (lightgray)},
      shader=faceted,
      samples = 17,
      domain=-0.8:0,
      domain y= 0 : pi/2,
      opacity = 1
    ]
    ({sqrt(1-x^2) * cos(deg(y))},
     {sqrt( 1-x^2 ) * sin(deg(y))},
     {-x});

    \addplot3[
      black,
      mark = none,
      very thick,
      domain=0:pi/4,
      samples y = 0
    ] 
    ({sqrt(1) * cos(deg(x))},
     {sqrt(1) * sin(deg(x))},
     {0});
    
    \addplot3[
      black,
      mark = none,
      very thick,
      domain=0:-.6,
      samples y = 0
    ] 
    ({sqrt(1-x^2) * cos(deg(0))},
     {sqrt( 1-x^2 ) * sin(deg(0))},
     {-x});
     
    \addplot3[
      black,
      mark = none,
      domain=0:pi/4,
      samples y = 0
    ] 
    ({sqrt(1-.60^2) * cos(deg(x))},
     {sqrt(1-.60^2 ) * sin(deg(x))},
     {0.60});
     
    \addplot3[
      black,
      mark = none,
      domain=0:-.6,
      samples y = 0
    ] 
    ({sqrt(1-x^2) * cos(deg(pi/4))},
     {sqrt( 1-x^2 ) * sin(deg(pi/4))},
     {-x});
    
    \node[circle, fill, minimum size = 4pt] (b0) at (axis cs: 0.566, 0.566, 0.6){};
    
    \node[anchor = west] (b1) at (axis cs: 0.3, 0.7, 0.6){};
    \node[anchor = west,left = 0.18cm,below = 0.1cm,fill=white,rounded corners=2pt] (b1l) at (b1){\scalebox{0.7}{$\boldsymbol{e}_1^b$}};
    
    \node[anchor = south,fill=white,rounded corners=2pt] (b2) at (axis cs: .5, .3, 0.6){\scalebox{0.7}{$\boldsymbol{e}_2^b$}};
    
    \node[anchor = center] (xpu) at (axis cs: 0.066, 1.066, 0.6){};
    \node[anchor = west, right = 0.15cm](xps) at (xpu){\scalebox{0.7}{$\boldsymbol{x}^p_s$}};
    
    \node[anchor = west,fill=white,rounded corners=2pt] (xpv) at (axis cs: 0.266, 0.266, 0.85) {\scalebox{0.7}{$\boldsymbol{x}^p_y$}};
    
    \node(en) at (axis cs: .65, .6, 1.5){};
    \node(enl)[anchor = east, right = 0.1cm,fill=white,rounded corners=2pt] at (en){\scalebox{0.7}{$\boldsymbol{e}_n^p,\boldsymbol{e}_3^b$}};
    
    \draw[thick,blue,->](b0)--(b1);
    \draw[thick,blue,->](b0)--(b2);
    
    \draw[thick,black,->](b0)--(xpu);
    \draw[thick,black,->](b0)--(xpv);
    \draw[postaction={draw,black,dash pattern= on 3pt off 3pt,dash phase=3pt,thick,->}][blue,dash pattern= on 3pt off 3pt,thick,->](b0)--(en);
    
    \node[] (thu1) at (xpu){};
    \node[] (thu2) at (axis cs: 0,0.85,0.7){};
    \draw[very thin,] (b0) -- (thu1);
    \draw[very thin,] (b0) -- (thu2);
    \draw[bend right, ->] (thu1) to (thu2) ;
    \node (thu) at (axis cs: 0, 1, .9) {\scalebox{0.7}{$\theta^s$}};
    \node[coordinate, pin = {[pin distance=0.5cm,pin edge = {thick,dashed,black}]left:{\scalebox{0.7}{Constraint Surface }}}] at (axis cs: .588, 0.119, 0.8){};
    \node[coordinate, pin = {[pin distance=0.5cm,pin edge = {thick,dashed,black}]left:{\scalebox{0.7}{Road Surface }}}] at (axis cs: .6, 0, 0.46){};
    \node[coordinate, pin = {[pin distance=1.4cm, pin edge = {thick,dashed,black}]0:{\scalebox{0.7}{ $\boldsymbol{x}^p$ and $\boldsymbol{x}^g$ (COM)}}}] at (b0){};

    \node (A)  at (axis cs: 1, 0,0){};
    \node[below = 0.4cm, left = 0.4cm] (Ay) at (A){};
    \node[below = 0.2cm, left = 0.2cm] (Ayh) at (A){};
    \node (B)  at (axis cs: 0.8, 0,0.6){};
    \node[below = 0.4cm, left = 0.4cm] (By) at (B){};
    \node[below = 0.2cm, left = 0.2cm] (Byh) at (B){};
    \draw[very thin] (A) -- (Ay);
    \draw[very thin] (B) -- (By);
    \draw[bend right=12, ->] (Ayh) to (Byh);
    \node[fill=white,rounded corners=2pt, inner sep = 2pt] (y) at (axis cs: 0.9, -0.1,0.4){\scalebox{0.7}{$y$}};
    
    \node (As) at (axis cs: 1.2,-0,0){};
    \node (Ash) at (axis cs: 1.1,-0,0){};
    \node (C)  at (axis cs: 0.707,0.707,0){};
    \node (Cs)  at (axis cs: .85,.85,0){};
    \node (Csh)  at (axis cs: 0.77,0.77,0.0){};
    \draw[very thin] (A) -- (As);
    \draw[very thin] (C) -- (Cs);
    \draw[bend right, ->] (Ash) to (Csh);
    \node[fill=white,rounded corners=2pt, inner sep = 2pt] (s) at (axis cs: 1.025, 0.40,0){\scalebox{0.7}{$s$}};

    \end{axis}

    \end{tikzpicture}
        
    \caption{Coordinate systems and road and constraint surface. Superscript $g$ denotes vectors in the global frame, $b$ for the body frame and $p$ for the parametric frame. $\theta^s$ is an angle between the vehicle body and the partial derivative $\boldsymbol{x}_s^p$}
    \label{fig:frames_of_reference}
\end{figure}

We focus on vehicle motion planning models with a curvilinear state, such as those in~\cite{MICAELLI1994249}. These models define position and orientation relative to a centerline and are a natural choice for path planning and control~\cite{6225063,frenet_tracking_paper}. However, classical 3D vehicle models such as the one reported in~\cite[Ch.~9]{guiggiani_book} describe vehicle pose relative to a fixed Euclidean frame of reference, limiting path planning and control to planar surfaces. The work in {{\cite{3d_part_1,3d_part_2}} explores vehicle dynamics on ribbon-shaped roads; in our approach we derive equations of motion for a vehicle driving on a generic parametric surface where the pose of the vehicle is described by the surface parameterization in the form of a ``parametric pose", which is formally defined in section {\ref{sec:prelim_parametric_surfaces}}}. 
We make four key modeling assumptions common for motion planning and control:
\begin{enumerate}
    \item The entire vehicle can be treated as a single rigid body; we do not model suspensions.
    \item The equations of motion are written about the vehicle's center of mass (COM).
    \item Out of plane road curvature is small relative to the length of the vehicle. 
    \item Contact with the road is never lost. 
\end{enumerate}
In Section \ref{sec:Conclusion} we remark on relaxing Assumption 1 by including suspensions in the proposed approach. Assumptions 3 and 4 imply that there exists a 2D surface in 3D that always contains the rigid body vehicle center of mass such that the surface normal vector is aligned with the vehicle body normal vector. We refer to this constraint as ``tangent contact" and the surface as the ``constraint surface"; this is illustrated in Figure \ref{fig:frames_of_reference}.  

The contributions of this paper can be summarized as follows:
\begin{itemize}
    \item We derive equations of motion for a rigid body in tangent contact with a parametric surface where the pose of the body is described by the surface parameterization.
    \item We demonstrate our approach generalizes vehicle models that reference a centerline, such as the Frenet model.
    \item We derive a kinematic model of a vehicle driving on a nonplanar parametric surface.
    \item We use the nonplanar kinematic model in a MPC controller and show it outperforms existing controllers, including MPC based on planar vehicle models. 
\end{itemize}

This paper is structured as follows: In section \ref{sec:preliminaries} we introduce background material. In section \ref{sec:Motion Along a Surface} we derive equations of motion for a rigid body in tangent contact with a parametric surface. We then apply these equations to specific surface parameterizations in section \ref{sec:surface_parameterizations} and derive a corresponding kinematic vehicle model in section \ref{sec:Vehicle_Models}. In section \ref{sec:control} we introduce a MPC algorithm for controlling a vehicle on a nonplanar surface which we simulate and test in sections \ref{sec:Simulation} and \ref{sec:Results}.

%%%%%%%%%%%%%%%%%%%%%%%%%%%%%%%%%%%%%%%%%%%%%%%%%%%%%%%%%%

\section{Preliminaries} \label{sec:preliminaries}

\aftergroup\restoregeometry

%%%%%%%%%%%%%%%%%%%%%%%%%%%%%%%%%%%%%%%%%%%%%%%%
\subsection{Frames of Reference}
We use two frames of reference. The first, an inertial frame of reference, is one in which the surface the vehicle is driving on is stationary. The second, a noninertial frame, is one in which the vehicle is stationary with its center of mass at the origin. The inertial frame is referred to as the global frame and the noninertial frame as the body frame. We distinguish between the two with superscript $g$ and $b$ respectively. Components of body frame vectors are indexed by $\left\{1,2,3\right\}$. 

For the global frame we use the orthonormal basis $\left\{\boldsymbol{e}_1^g, \boldsymbol{e}_2^g, \boldsymbol{e}_3^g\right\}$ and for the body frame the orthonormal basis $\left\{ \boldsymbol{e}_1^b, \boldsymbol{e}_2^b, \boldsymbol{e}_3^b\right\}$. Both are shown in Figure \ref{fig:frames_of_reference}. We use the ISO body frame basis from~\cite{ISO8855} such that $\boldsymbol{e}_1^b$ points in the direction the vehicle is facing, $\boldsymbol{e}_2^b$ points to the left of the vehicle and $\boldsymbol{e}_3^b$ is the normal direction of the vehicle. 

We define $\boldsymbol{x}$ as the position of the center of mass relative to the global frame origin, $\boldsymbol{v}$ the linear velocity of the vehicle and $\boldsymbol{\omega}$ the angular velocity of the vehicle. These are vectors which may be expressed in any basis. We place a superscript above the vector to indicate when it is expressed in a particular frame, $g$ for the global frame and $b$ for the body frame. For instance, $\boldsymbol{v}^b = v_1^b \boldsymbol{e}_1^b + v_2^b \boldsymbol{e}_2^b + v_3^b \boldsymbol{e}_3^b$. Furthermore, we use $\frac{d}{dt}$ to denote the time derivative of a vector from the perspective of the global frame and $\dot{\ }$ to denote the component-wise time derivative of a vector. These differ for vectors expressed in the body frame, where the time derivatives of the rotating basis vectors are 
\begin{subequations}
    \begin{equation}
        \frac{d}{dt}\boldsymbol{e}_1^b = \omega_3^b \boldsymbol{e}_2^b - \omega_2^b \boldsymbol{e}_3^b 
    \end{equation}
    \begin{equation}
        \frac{d}{dt}\boldsymbol{e}_2^b = \omega_1^b \boldsymbol{e}_3^b - \omega_3^b \boldsymbol{e}_1^b 
    \end{equation}
    \begin{equation}
        \frac{d}{dt}\boldsymbol{e}_3^b = \omega_2^b \boldsymbol{e}_1^b - \omega_1^b \boldsymbol{e}_2^b.
    \end{equation}
\end{subequations}
As a result, $\dot{\boldsymbol{v}}^b = \dot{v}_1^b \boldsymbol{e}_1^b + \dot{v}_2^b \boldsymbol{e}_2^b + \dot{v}_3^b \boldsymbol{e}_3^b$ while $\frac{d}{dt} \boldsymbol{v}^b = \dot{\boldsymbol{v}}^b + v_1^b \frac{d}{dt}\boldsymbol{e}_1^b + v_2^b \frac{d}{dt}\boldsymbol{e}_2^b + v_3^b \frac{d}{dt}\boldsymbol{e}_3^b$.

%%%%%%%%%%%%%%%%%%%%%%%%%%%%%%%%%%%%%%%%%%%%%%%%
\subsection{Parametric Surfaces} \label{sec:prelim_parametric_surfaces}
We represent the constraint surface as the parametric surface:
\begin{equation}
    \boldsymbol{x}^p(s,y) = x^p_1(s,y) \boldsymbol{e}^g_1 + x^p_2(s,y) \boldsymbol{e}^g_2 + x^p_3(s,y) \boldsymbol{e}^g_3.
\end{equation}
It is often useful to define $y$ as the distance from the center of a lane and $s$ the path length of the lane (see Figure \ref{fig:frames_of_reference}); we leave them arbitrary for now to encompass other surfaces, such as elevation maps. We use superscript $p$ to refer to variables and frames related to the constraint surface. We also use $\boldsymbol{x}^p$ as shorthand for $\boldsymbol{x}^p(s,y)$. We denote partial derivatives of the parametric surface using subscripts: $\boldsymbol{x}^p_s = \frac{\partial}{\partial s} \boldsymbol{x}^p$ and $\boldsymbol{x}^p_y= \frac{\partial}{\partial y} \boldsymbol{x}^p$, which are shown in Figure \ref{fig:frames_of_reference}. {These define the surface normal vector:}
\begin{equation} \label{eq:normal_vector}
    \boldsymbol{e}^p_n = \frac{\boldsymbol{x}^p_s \times \boldsymbol{x}^p_y }{||\boldsymbol{x}^p_s \times \boldsymbol{x}^p_y ||}. 
\end{equation}
Since the components of $\boldsymbol{x}^p$ are functions of $s$ and $y$, $\frac{d}{dt}\boldsymbol{x}^p$ must be found using the chain rule:
\begin{equation} \label{eq:parametric_time_derivative}
    \frac{d}{dt}\boldsymbol{x}^p = \boldsymbol{x}^p_s \dot{s} + \boldsymbol{x}^p_y \dot{y}.
\end{equation}
Our definition of tangent contact implies that the vehicle COM coincides with the parametric surface and the surface normal vector $(\boldsymbol{e}^p_n)$ is aligned with the vehicle body normal vector. This corresponds to:
\begin{subequations} \label{eq:tangent_contact_simple}
    \begin{equation}
        \boldsymbol{x}^p = \boldsymbol{x}^g
    \end{equation}
    \begin{equation}
        \boldsymbol{e}^p_n = \boldsymbol{e}^b_3.
    \end{equation}
\end{subequations}
We assume that the surface is regular, which means $\boldsymbol{x}_s^p$ and $\boldsymbol{x}_y^p$, the tangent vectors of the surface, are nonzero and linearly independent~\cite[Ch.~2]{differential_geometry_of_curves_and_surfaces}. Note that for complex surfaces $\boldsymbol{x}^p_s$ and $\boldsymbol{x}^p_y$ may not be normal or orthogonal. We also assume that this surface is time invariant and smooth. 

Regularity allows us to rewrite \eqref{eq:tangent_contact_simple} as
\begin{subequations} \label{eq:tangent_contact_constraints}
    \begin{equation} \label{eq:surface_constraint}
        \boldsymbol{x}^p = \boldsymbol{x}^g
    \end{equation}
    \begin{equation} \label{eq:tangent_constraint_1}
        \boldsymbol{e}_3^b \cdot \boldsymbol{x}^p_s = 0
    \end{equation}
    \begin{equation} \label{eq:tangent_constraint_2}
        \boldsymbol{e}_3^b \cdot \boldsymbol{x}^p_y = 0.
    \end{equation}
\end{subequations}

 Imposing \eqref{eq:tangent_contact_constraints} on an unconstrained rigid body allows $s$ and $y$ to fully describe the position of the rigid body but leaves one rotational degree of freedom. To fully describe the pose of a vehicle body using a parametric surface we introduce a parametric orientation using the angle $\theta^s$ between $\boldsymbol{e}_1^b$ and $\boldsymbol{x}_s^p$ depicted in Figure \ref{fig:frames_of_reference} and defined below:
\begin{subequations}\label{eq:parametric_angle}
\begin{equation}\label{eq:parametric_angle_a}
    \cos(\theta^s) = \frac{\boldsymbol{e}_1^b  \cdot \boldsymbol{x}^p_s}{||\boldsymbol{x}^p_s||}
\end{equation}
\begin{equation}
    \sin(\theta^s) = \frac{-\boldsymbol{e}_2^b \cdot \boldsymbol{x}^p_s}{||\boldsymbol{x}^p_s||}.
\end{equation}
\end{subequations}
Here $||\ ||$ denotes the $L^2$ norm and $\boldsymbol{a}\cdot \boldsymbol{b}$ the dot product. Other angles may be used by choosing other pairs of vectors. We define $\left(s,y, \theta^s\right)$ as the ``parametric pose" of a vehicle. The parametric pose fully describes the position and orientation of a rigid body in tangent contact with the constraint surface and requires fewer equations of motion than global frame position and orientation. We derive these equations of motion in Section \ref{sec:Motion Along a Surface}.

%%%%%%%%%%%%%%%%%%%%%%%%%%%%%%%%%%%%%%%%%%%%%%%%
\subsection{Rigid Body Dynamics}
The Newton-Euler equations of motion for an unconstrained rigid body in 3D are~\cite[p.~115]{mechanics_landau_lifshitz}:

\begin{subequations}\label{eq:equations_of_motion}
    \begin{equation}
        m\left(\dot{\boldsymbol{v}}^b + \boldsymbol{\omega}^b \times \boldsymbol{v}^b \right) = \boldsymbol{F}^b
    \end{equation}
    \begin{equation}\label{eq:cons_ang_momentum_full}
        \left(\boldsymbol{I}^b\dot{\boldsymbol{\omega}}^b + \boldsymbol{\omega}^b \times \left(\boldsymbol{I}^b \boldsymbol{\omega}^b \right) \right) = \boldsymbol{K}^b,
    \end{equation}
\end{subequations}

where $m$ and $\boldsymbol{I}^b$ are the mass and moment of inertia of the body and $\boldsymbol{I}^b$ is expressed in the body frame basis. $\boldsymbol{F}^b$ and $\boldsymbol{K}^b$ are net force and torque on the body with components in the body frame basis. 

In the body frame the moment of inertia tensor is time invariant, simplifying integration of \eqref{eq:equations_of_motion}. For most vehicles the inertia tensor is also diagonal~\cite[Ch.~9]{guiggiani_book}  with elements $I_1^b$ through $I_3^b$, which we assume is the case moving forwards. Next we impose constraints on \eqref{eq:equations_of_motion} for a rigid body in tangent contact with a parametric surface. 

%%%%%%%%%%%%%%%%%%%%%%%%%%%%%%%%%%%%%%%%%%%%%%%%
\section{Motion Along a Surface} \label{sec:Motion Along a Surface}
In this section we simplify rigid body motion \eqref{eq:equations_of_motion} when constrained by \eqref{eq:tangent_contact_constraints}. We also describe rigid body motion using ${s}$, ${y}$ and ${\theta}^s$ instead of global frame position and orientation. We achieve both by differentiating the constraints \eqref{eq:tangent_contact_constraints} with respect to time in the global frame.

%%%%%%%%%%%%%%%%%%%%%%%%%%%%%%%%%%%%%%%%%%%%%%%%
\subsection{Linear Velocity Constraint}
The time derivative of global frame position in \eqref{eq:surface_constraint} is velocity, which we express in the body frame. Using the time derivative of $\boldsymbol{x}^p$ from \eqref{eq:parametric_time_derivative}, \eqref{eq:surface_constraint} can be rewritten as:
\begin{equation}\label{eq:parametric_velocity_unsimplified}
    \dot{s}\boldsymbol{x}^p_s + \dot{y}\boldsymbol{x}^p_y = v^b_1 \boldsymbol{e}^b_1 + v^b_2 \boldsymbol{e}^b_2 + v^b_3 \boldsymbol{e}^b_3.
\end{equation}
Equations \eqref{eq:tangent_contact_constraints} and \eqref{eq:parametric_velocity_unsimplified} result in:
\begin{equation}\label{eq:no_normal_motion}
    v_3^b = 0.
\end{equation}
We can simplify the remainder of \eqref{eq:parametric_velocity_unsimplified} using dot products with the vectors $\boldsymbol{x}^p_s$ and $\boldsymbol{x}^p_y$:

\begin{equation} \label{eq:parametric_velocity}
    \begin{split}
        \underbrace{
        \begin{bmatrix}
        \boldsymbol{x}^p_s \cdot \boldsymbol{e}_1^b & \boldsymbol{x}^p_s \cdot \boldsymbol{e}_2^b\\
        \boldsymbol{x}^p_y \cdot \boldsymbol{e}_1^b & \boldsymbol{x}^p_y \cdot \boldsymbol{e}_2^b
        \end{bmatrix}
        }_{\mathbf{J}}
        \begin{bmatrix}
        v_1^b \\
        v_2^b
        \end{bmatrix}
        =
        \underbrace{
        \begin{bmatrix}
        \boldsymbol{x}^p_s \cdot \boldsymbol{x}^p_s  & \boldsymbol{x}^p_s \cdot \boldsymbol{x}^p_y \\
        \boldsymbol{x}^p_s \cdot \boldsymbol{x}^p_y & \boldsymbol{x}^p_y \cdot \boldsymbol{x}^p_y
        \end{bmatrix}
        }_{\mathbf{I}}
        \begin{bmatrix}
        \dot{s}\\
        \dot{y}
        \end{bmatrix}.
    \end{split}
\end{equation}
The matrix on the right hand side of \eqref{eq:parametric_velocity} is known as the first fundamental form\footnote{Regrettably it is common practice to use ``I" for both the first fundamental form and moment of inertia. From here on the moment of inertia will always appear in diagonal components $(I_1^b, I_2^b, I_3^b)$ and the first fundamental form as $(\mathbf{I})$ or expanded as a matrix.} $(\mathbf{I})$ of a parametric surface~\cite[Ch.~2]{differential_geometry_of_curves_and_surfaces}. We use the symbol $(\mathbf{J})$ for the Jacobian matrix of $\boldsymbol{x}^p$ on the left hand side. It follows that:
\begin{equation} \label{eq:parametric_velocity_simplified}
    \begin{bmatrix}
    \dot{s}\\
    \dot{y}
    \end{bmatrix}
    =
    \mathbf{I}^{-1}
    \mathbf{J}
    \begin{bmatrix}
        v_1^b \\
        v_2^b
    \end{bmatrix}.
\end{equation}
For a regular surface $\mathbf{I}$ is always invertible and when $\mathbf{I}$ is invertible a surface is regular \cite[Ch.~2]{differential_geometry_of_curves_and_surfaces}.

%%%%%%%%%%%%%%%%%%%%%%%%%%%%%%%%%%%%%%%%%%%%%%%%
\subsection{Angular Velocity Constraint}
The time derivatives of \eqref{eq:tangent_constraint_1} and \eqref{eq:tangent_constraint_2} can be found using the time derivative of the dot product:

\begin{subequations}
    \begin{equation}
        \frac{d}{dt}\left(\boldsymbol{x}^p_s\right) \cdot \boldsymbol{e}_n^p + \boldsymbol{x}_s^p \cdot \frac{d}{dt} \left( \boldsymbol{e}_3^b\right)= 0
    \end{equation}
    \begin{equation}
        \frac{d}{dt}\left(\boldsymbol{x}^p_y\right) \cdot \boldsymbol{e}_n^p + \boldsymbol{x}_y^p \cdot \frac{d}{dt} \left(\boldsymbol{e}_3^b\right) = 0,
    \end{equation}
\end{subequations}
and then expanding the time derivatives of $\boldsymbol{x}^p_s$, $\boldsymbol{x}^p_y$ and $\boldsymbol{e}^b_3$:
\begin{equation} \label{eq:surface_angular_velocity_unsimplified}
    \begin{split}
        \underbrace{
        \begin{bmatrix}
        \boldsymbol{x}^p_{ss} \cdot \boldsymbol{e}_n^p & \boldsymbol{x}^p_{sy} \cdot \boldsymbol{e}_n^p \\
        \boldsymbol{x}^p_{ys} \cdot \boldsymbol{e}_n^p & \boldsymbol{x}^p_{yy} \cdot \boldsymbol{e}_n^p
        \end{bmatrix}
        }_{\mathbf{II}}
        \begin{bmatrix}
        \dot{s} \\ \dot{y}
        \end{bmatrix}
        =
        \begin{bmatrix}
        \boldsymbol{x}^p_s \cdot \boldsymbol{e}_1^b & \boldsymbol{x}^p_s \cdot \boldsymbol{e}_2^b \\
        \boldsymbol{x}^p_y \cdot \boldsymbol{e}_1^b & \boldsymbol{x}^p_y \cdot \boldsymbol{e}_2^b
        \end{bmatrix}
        \begin{bmatrix}
        -\omega_2^b \\ \omega_1^b
        \end{bmatrix}.
    \end{split}
\end{equation}
The matrix on the left hand side of \eqref{eq:surface_angular_velocity_unsimplified} is known as the second fundamental form of a parametric surface~\cite[Ch.~3]{differential_geometry_of_curves_and_surfaces}, which we denote by $\mathbf{II}$. Altogether, we have the following relationship for $\omega_1^b$ and $\omega_2^b$:

\begin{equation} \label{eq:surface_angular_velocity}
    \begin{bmatrix}
        -\omega_2^b \\ \omega_1^b
    \end{bmatrix}
    =\mathbf{J}^{-1} \mathbf{II}\ \mathbf{I}^{-1} \mathbf{J}
    \begin{bmatrix}
    v_1^b \\
    v_2^b
    \end{bmatrix}.
\end{equation}

This allows us to remove $\omega_1^b$ and $\omega_2^b$ from \eqref{eq:equations_of_motion} and use $v_1^b$ and $v_2^b$ instead.

%%%%%%%%%%%%%%%%%%%%%%%%%%%%%%%%%%%%%%%%%%%%%%%%
\subsection{Parametric Angular Velocity}
Finally we take the time derivative of \eqref{eq:parametric_angle_a} to derive an equation of motion for $\theta^s$:
\begin{equation} \label{eq:track_angular_velocity_unsimplified}
    \begin{split}
        -\sin(\theta^s)\dot{\theta}^s ||\boldsymbol{x}^p_s||^2 = 
        \\
        \left(\frac{d}{dt}\boldsymbol{e}_1^b \cdot \boldsymbol{x}^p_s + \boldsymbol{e}_1^b \cdot \dot{\boldsymbol{x}^p_s} \right) ||\boldsymbol{x}^p_s|| - \boldsymbol{e}_1^b \cdot \boldsymbol{x}^p_s \frac{d}{d t} ||\boldsymbol{x}^p_s||,
    \end{split}
\end{equation}
into which we can substitute the equation for $\sin(\theta^s)$ and expand the time derivatives:
\begin{equation}\label{eq:theta_s_nastiest}
    \begin{split}
        \dot{\theta}^s = \frac{||\boldsymbol{x}^p_s||}{\boldsymbol{e}_2^b \cdot \boldsymbol{x}^p_s} \frac{\omega_3 \boldsymbol{e}_2^b \cdot \boldsymbol{x}^p_s + \boldsymbol{e}_1^b \cdot \boldsymbol{x}^p_{ss} \dot{s} + \boldsymbol{e}_1^b \cdot \boldsymbol{x}^p_{sy} \dot{y}}{||\boldsymbol{x}^p_s||} 
        \\
        - \frac{||\boldsymbol{x}^p_s||}{\boldsymbol{e}_2^b \cdot \boldsymbol{x}^p_s} \frac{(\boldsymbol{e}_1^b \cdot \boldsymbol{x}^p_s)(\boldsymbol{x}^p_s \cdot \boldsymbol{x}^p_s)^{-1/2} (\dot{\boldsymbol{x}^p_s}\cdot \boldsymbol{x}^p_s)}{\boldsymbol{x}^p_s \cdot \boldsymbol{x}^p_s}.
    \end{split}
\end{equation}
Factoring out terms,
\begin{equation}\label{eq:theta_s_nasty}
    \begin{split}
        \dot{\theta}^s = \omega_3^b + \frac{\boldsymbol{e}_1^b \cdot \boldsymbol{x}^p_{ss} -(\boldsymbol{e}_1^b\cdot \boldsymbol{x}^p_s)(\boldsymbol{x}^p_s\cdot \boldsymbol{x}^p_{ss})(\boldsymbol{x}^p_s\cdot \boldsymbol{x}^p_s)^{-1}}{\boldsymbol{e}_2^b \cdot \boldsymbol{x}^p_s}\dot{s}
        \\
        +\frac{\boldsymbol{e}_1^b \cdot \boldsymbol{x}^p_{sy} -(\boldsymbol{e}_1^b\cdot \boldsymbol{x}^p_s)(\boldsymbol{x}^p_s\cdot \boldsymbol{x}^p_{sy})(\boldsymbol{x}^p_s\cdot \boldsymbol{x}^p_s)^{-1}}{\boldsymbol{e}_2^b \cdot \boldsymbol{x}^p_s}\dot{y}.
    \end{split}
\end{equation}
Equation \eqref{eq:theta_s_nasty} is not useful since the denominator can be zero. However, if we expand $\boldsymbol{x}^p_{s}$, $\boldsymbol{x}^p_{ss}$ and $\boldsymbol{x}^p_{sy}$ into their $\boldsymbol{e}_1^b$, $\boldsymbol{e}_2^b$ and $\boldsymbol{e}_3^b$ components ({see Appendix {\ref{app:theta_s}}}) we obtain the expression
\begin{equation}\label{eq:theta_s_unsimplified}
    \begin{split}
        \dot{\theta}^s = \omega_3^b + \frac{(\boldsymbol{e}_1^b\cdot \boldsymbol{x}^p_{ss})(\boldsymbol{e}_2^b\cdot \boldsymbol{x}^p_s) - (\boldsymbol{e}_2^b \cdot \boldsymbol{x}^p_{ss})(\boldsymbol{e}_1^b\cdot \boldsymbol{x}^p_s)}{\boldsymbol{x}^p_s\cdot \boldsymbol{x}^p_s}\dot{s}
        \\
        +\frac{(\boldsymbol{e}_1^b \cdot \boldsymbol{x}^p_{sy})(\boldsymbol{e}_2^b\cdot \boldsymbol{x}^p_s) - (\boldsymbol{e}_2^b \cdot \boldsymbol{x}^p_{sy})(\boldsymbol{e}_1^b\cdot \boldsymbol{x}^p_s)}{\boldsymbol{x}^p_s\cdot \boldsymbol{x}^p_s}\dot{y}.
    \end{split}
\end{equation}
Equation \eqref{eq:theta_s_unsimplified} can be rewritten with the cross product:
\begin{equation} \label{eq:track_angular_velocity}
    \begin{split}
        \dot{\theta}^s = \omega_3^b + 
          \frac{\left(\boldsymbol{x}^p_{ss}\times \boldsymbol{x}^p_s\right)\cdot \boldsymbol{e}_n^p}{\boldsymbol{x}^p_s \cdot \boldsymbol{x}^p_s}\dot{s} 
        \\
        + \frac{\left(\boldsymbol{x}^p_{sy}\times \boldsymbol{x}^p_s\right)\cdot \boldsymbol{e}_n^p}{\boldsymbol{x}^p_s \cdot \boldsymbol{x}^p_s}\dot{y}.
    \end{split}
\end{equation}
Equations \eqref{eq:track_angular_velocity_unsimplified} through \eqref{eq:track_angular_velocity} are equivalent however we use \eqref{eq:track_angular_velocity} since the cross product terms are independent of body orientation.

%%%%%%%%%%%%%%%%%%%%%%%%%%%%%%%%%%%%%%%%%%%%%%%%
\subsection{Constrained Equations of Motion} \label{sec:constrained_equations_of_motion}

We simplify \eqref{eq:equations_of_motion} in tangent contact with a parametric surface using \eqref{eq:no_normal_motion}, \eqref{eq:parametric_velocity_simplified}, \eqref{eq:surface_angular_velocity} and \eqref{eq:track_angular_velocity}. For clarity we split the equations of motion into dynamic equations \eqref{eq:constrained_eom_dynamic}, kinematic equations \eqref{eq:constrained_eom_parametric_kinematic} and constraints \eqref{eq:constrained_eom_constraints}:

\begin{subequations} \label{eq:constrained_eom_dynamic}
    \begin{equation}
        \dot{v}^b_1 = \omega^b_3 v^b_2 + F_1^b/m
    \end{equation}
    \begin{equation}
        \dot{v}^b_2 = -\omega^b_3 v^b_1 + F_2^b/m
    \end{equation}
    \begin{equation}
        \dot{\omega}^b_3 = ((I^b_1-I^b_2)\omega^b_1\omega^b_2 + K^b_3)/I^b_3 
    \end{equation}
\end{subequations}

\begin{subequations} \label{eq:constrained_eom_parametric_kinematic}
    \begin{equation}
        \begin{bmatrix}
        \dot{s}\\
        \dot{y}
        \end{bmatrix}
        =
        \mathbf{I} ^{-1}
        \mathbf{J}
        \begin{bmatrix}
            v_1^b \\
            v_2^b
        \end{bmatrix}
    \end{equation}
    \begin{equation}\label{eq:dthetau}
            \begin{split}
            \dot{\theta}^s = \omega_3^b + 
              \frac{\left(\boldsymbol{x}^p_{ss}\times \boldsymbol{x}^p_s\right)\cdot \boldsymbol{e}_n^p}{\boldsymbol{x}^p_s \cdot \boldsymbol{x}^p_s}\dot{s} 
            \\
            + \frac{\left(\boldsymbol{x}^p_{sy}\times \boldsymbol{x}^p_s\right)\cdot \boldsymbol{e}_n^p}{\boldsymbol{x}^p_s \cdot \boldsymbol{x}^p_s}\dot{y}
        \end{split}
    \end{equation}
\end{subequations}

\begin{subequations} \label{eq:constrained_eom_constraints}
    \begin{equation} \label{eq:shape_operator}
        \begin{bmatrix}
            -\omega_2^b \\ \omega_1^b
        \end{bmatrix}
        =\mathbf{J}^{-1} \mathbf{II}\ \mathbf{I}^{-1} \mathbf{J}
        \begin{bmatrix}
        v_1^b \\
        v_2^b
        \end{bmatrix}
    \end{equation}
    \begin{equation} \label{eq:v3_constraint}
        v_3^b = 0
    \end{equation}
    \begin{equation} \label{eq:F3_constraint}
        \dot{v}_3^b=0 = -\omega_1^b v_2^b + \omega_2^b v_1^b + F_3^b/m
    \end{equation}
    \begin{equation} \label{eq:K1_constraint}
        \dot{\omega}_1^b = ((I^b_2-I^b_3)\omega^b_2\omega^b_3 + K_1^b)/I_1^b
    \end{equation}
    \begin{equation} \label{eq:K2_constraint}
        \dot{\omega}_2^b = ((I_3^b-I_1^b)\omega^b_3\omega^b_1 + K_2^b)/I_2^b.
    \end{equation}
\end{subequations}
As a result, we can model a vehicle on a parametric surface using the parametric pose of \eqref{eq:constrained_eom_parametric_kinematic} rather than global position and orientation. This eliminates unnecessary variables and allows $s$ and $y$ to hold specific meaning, for instance $s$ may be path length of a lane and $y$ distance from its centerline. This is common for model-based control for vehicle navigation, but we remark that no specific physical meaning is required for this derivation. Modeling a physical vehicle requires equations for $F_1^b$, $F_2^b$ and $K_3^b$ or kinematic steering constraints. We introduce the latter and remark on the former in Section \ref{sec:Vehicle_Models}. Note that one can substitute \eqref{eq:shape_operator} into \eqref{eq:constrained_eom_dynamic}; we avoid this for readability. Furthermore, equations (\ref{eq:v3_constraint}-\ref{eq:K2_constraint}) impact the weight distribution of a vehicle. 

%%%%%%%%%%%%%%%%%%%%%%%%%%%%%%%%%%%%%%%%%%%%%%%%
\subsection{Second Order Constraints}
Higher order constraints, ie. relating $\dot{\omega}_1^b$ to vehicle speed and acceleration, can be derived by differentiating \eqref{eq:parametric_velocity_simplified} and \eqref{eq:surface_angular_velocity} once more. This constrains net torque components through \eqref{eq:K1_constraint} and \eqref{eq:K2_constraint}, which alters the weight distribution of a vehicle. These constraints can be derived in a similar manner as the previous section but are unnecessary for the kinematic model developed in Section \ref{sec:Vehicle_Models}. 

%%%%%%%%%%%%%%%%%%%%%%%%%%%%%%%%%%%%%%%%%%%%
\section{Surface Parameterizations} \label{sec:surface_parameterizations}
In this section we focus on parametric surfaces parameterized by a curved centerline and linear offset\footnote{For fixed $s$, the cross section of the surface along $y$ is a line} from the centerline. {Assuming that the orthonormal tangent {($\boldsymbol{e}_s^c$)}, binormal {($\boldsymbol{e}_y^c$)} and normal {($\boldsymbol{e}_n^c$)} vectors of a centerline are given, the centerline position {$\boldsymbol{x}^c$} in the global frame is defined by:}
\begin{equation} \label{eq:general_centerline_ode}
    \frac{d \boldsymbol{x}^c}{d s} = \boldsymbol{e}_s^c(s).
\end{equation}
{We add a linear offset {$y\boldsymbol{e}^c_y$}, obtaining the parametric surface}:
\begin{equation} \label{eq:general_centerline}
    \boldsymbol{x}^p(s,y) = y \boldsymbol{e}_y^c(s) + \boldsymbol{x}^c(s)
\end{equation}
{Note that due to torsion, $\boldsymbol{e}_n^p \neq \boldsymbol{e}^c_n$ when $y\neq 0$ so $\boldsymbol{e}_n^p$ must be found using {\eqref{eq:normal_vector}}.} We leave the initial condition of \eqref{eq:general_centerline_ode} arbitrary as it does not affect the equations of motion. Using \eqref{eq:general_centerline} we obtain the partial derivatives of $\boldsymbol{x}^p$:

\begin{subequations} \label{eq:centerline_partial_derivatives}
    \begin{equation}
        \boldsymbol{x}^p_s = \boldsymbol{e}_s^c + y\frac{d}{d s}(\boldsymbol{e}_y^c) 
    \end{equation}
    \begin{equation}
        \boldsymbol{x}^p_y = \boldsymbol{e}_y^c
    \end{equation}
    \begin{equation}
        \boldsymbol{x}^p_{ss} = \frac{d}{d s} \boldsymbol{e}_s^c + y\frac{d^2}{d s^2}(\boldsymbol{e}_y^c) 
    \end{equation}
    \begin{equation}
        \boldsymbol{x}^p_{sy} = \frac{d}{d s}\left(\boldsymbol{e}_y^c\right)
    \end{equation}
    \begin{equation}
        \boldsymbol{x}^p_{yy} = 0.
    \end{equation}
\end{subequations}
Since $\boldsymbol{e}_s^c$ and $\boldsymbol{e}_y^c$ are orthonormal, $\boldsymbol{x}^p_s \cdot \boldsymbol{x}^p_y = \boldsymbol{e}_s^c \cdot \boldsymbol{e}_y^c + y \frac{d}{ds}\left(\boldsymbol{e}_y^c\right) \cdot \boldsymbol{e}_y^c = 0$. This allows us to compute $\mathbf{J}$ directly from \eqref{eq:parametric_angle}:
\begin{equation} \label{eq:simplified_J}
    \mathbf{J} = 
    \begin{bmatrix}
        \cos\left(\theta^s\right) || \boldsymbol{x}^p_s|| & -\sin\left(\theta^s\right)  || \boldsymbol{x}^p_s|| \\
        \sin\left(\theta^s\right)  || \boldsymbol{x}^p_y|| & \cos\left(\theta^s\right)  || \boldsymbol{x}^p_y||
    \end{bmatrix}.
\end{equation}
In general, computing $\mathbf{J}$ also requires the angle between $\boldsymbol{x}^p_s$ and $\boldsymbol{x}^p_y$.

\subsection{Planar Frenet Frame Dynamics}
\begin{figure}
    \centering
    \begin{tikzpicture}

        \draw[thick, ->] (0,0)--(-1,1);
        \draw[thick, ->] (0,0)--(1,1);
        
        \node[] (x0) at (0.6,0) {$\boldsymbol{x}^c(s)$};
        \node[] (es) at (1.2,1) {$\boldsymbol{e}_s^c$};
        \node[] (ey) at (-0.7,1) {$\boldsymbol{e}_y^c$};
        
        \draw[red, ->] (0,0) arc (-45:0:3);
        \draw[red] (0,0) arc (-45:-90:3);
        
        \draw[thin] (0.1,-0.1) -- (.5,-.5);
        \draw[thin] (-2.12,-1.02) -- (-2.12,-1.586);
        
        \draw[thin, <-] (.3,-0.3) arc (-45:-60:3.424);
        \draw[thin, -] (-2.12,-1.32) arc (-90:-70:3.424);
        \node[] (s) at (-0.73,-1.0) {$s$};
        
        \draw[thin] (-0.1,-0.1) -- (-0.5,-0.5);
        \draw[thin] (-1.6,1.4) -- (-2.0, 1);
        \draw[thin] (-0.3,-0.3) -- (-0.85, 0.25);
        \draw[thin] (-1.25,0.65)   -- (-1.8, 1.2);
        \node[] (s) at (-1.05,0.45) {$y$};

        \node[thick,circle,inner sep=0pt, fill, minimum size = 0.1cm] (b0) at (-1.5,1.5){};
        \node[anchor = west] (x) at (-1.4, 1.5) {$\boldsymbol{x}^p(s,y)$};
        
        \draw[thin,dashed] (-1.5,1.5) -- (-0.5,2.5);
        \draw[thin,dashed] (-1.5,1.5) -- (-1.134, 2.866);
        \draw[thin,->] (-1,2) arc (45:75:0.707);
        \node[] (ths) at (-1,2.3) {$\theta^s$};
        
        \filldraw [fill=green, draw = black, fill opacity= 0.5, rotate around={-15:(-1.5,1.5)}] (-1.65,1.25) rectangle ++(0.3,0.5);
        \draw[blue, ->, rotate around={75:(-1.5,1.5)}] (-1.5,1.5) -- (-0.4,1.5);
        \draw[blue, ->, rotate around={165:(-1.5,1.5)}] (-1.5,1.5) -- (-0.8,1.5);
        
        \node[anchor = east] (v1) at (-1.2, 2.7) {$v_1^b$};
        \node[anchor = east] (v2) at (-2.1, 1.8) {$v_2^b$};
    \end{tikzpicture}
    \caption{Frenet Frame of Reference}
    \label{fig:Frenet_Frame}
\end{figure}
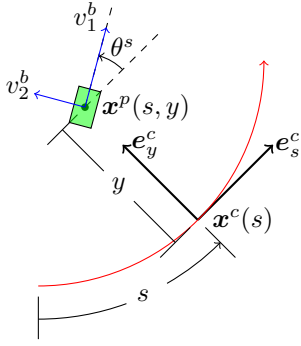

The Frenet frame of reference is the simplest surface of form \eqref{eq:general_centerline}. This is depicted in Figure \ref{fig:Frenet_Frame}. In the Frenet frame, $\boldsymbol{e}_y^c$ and $\boldsymbol{e}_s^c$ vary according to~\cite[Ch.~1]{differential_geometry_of_curves_and_surfaces}:
\begin{subequations} \label{eq:frenet_curvature}
    \begin{equation}
        \frac{d \boldsymbol{e}_y^c}{d s} = -\kappa \boldsymbol{e}_s^c
    \end{equation}
    \begin{equation}
        \frac{d \boldsymbol{e}_s^c}{d s} = \kappa \boldsymbol{e}_y^c,
    \end{equation}
\end{subequations}
where $\kappa$ is the curvature of the track\footnote{Some works instead use the radius of curvature $r = \sfrac{1}{\kappa}$, we use curvature since $r$ is infinite for a straight segment, prohibiting numerical interpolation of $r(s)$} and is a function of $s$. We assume arbitrary orthonormal initial conditions, as the initial orientation does not affect the equations of motion. A centerline is defined by $\kappa(s)$, and the surface is regular when $1-\kappa y > 0$\footnote{This will be obvious in the rank of the first fundamental form}; we assume this is always the case. The partial derivatives of $\boldsymbol{x}^p$ are:
\begin{subequations}
    \begin{equation}
        \boldsymbol{x}^p_s = \left(1-y\kappa\right) \boldsymbol{e}_s^c
    \end{equation}
    \begin{equation}
        \boldsymbol{x}^p_y = \boldsymbol{e}_y^c
    \end{equation}
    \begin{equation}
        \boldsymbol{x}^p_{ss} = -y\kappa_s \boldsymbol{e}_s^c +\left(1 -y\kappa\right) \kappa \boldsymbol{e}_y^c
    \end{equation}
    \begin{equation}
        \boldsymbol{x}^p_{sy} = -\kappa \boldsymbol{e}_s^c 
    \end{equation}
    \begin{equation}
        \boldsymbol{x}^p_{yy} = 0,
    \end{equation}
\end{subequations}
where $\kappa_s$ is the derivative of $\kappa$ with respect to s. Thus the first and second fundamental forms are 
\begin{equation}
    \mathbf{I} = 
    \begin{bmatrix}
    \left(1-\kappa y\right)^2 & 0 \\
    0 & 1
    \end{bmatrix},\ 
    \mathbf{II} = 
    \begin{bmatrix}
    0 & 0 \\ 0 & 0
    \end{bmatrix}.
\end{equation}
Since the second fundamental form is zero everywhere, $\omega_1^b = \omega_2^b = 0$. Equation \eqref{eq:dthetau} simplifies to:
\begin{equation}
    \dot{\theta}^s = \omega_3^b - \kappa \dot{s}.
\end{equation}
Using \eqref{eq:simplified_J}:
\begin{equation}
    \mathbf{J} = 
    \begin{bmatrix}
        (1-\kappa y)\cos(\theta^s) & -(1-\kappa y) \sin(\theta^s) \\
        \sin(\theta^s) & \cos(\theta^s)
    \end{bmatrix}.
\end{equation}

Now using \eqref{eq:constrained_eom_dynamic} through \eqref{eq:constrained_eom_constraints} we obtain the Frenet frame equations of motion:
\begin{subequations}
    \begin{equation}
        \dot{s} = \frac{v_1^b \cos(\theta^s) - v_2^b \sin(\theta^s)}{1-\kappa y}
    \end{equation}
    \begin{equation}
        \dot{y} = v_1^b \sin(\theta^s) + v_2^b \cos(\theta^s)
    \end{equation}
    \begin{equation}
        \dot{\theta}^s = \omega_3^b - \kappa \dot{s} 
    \end{equation}
    \begin{equation}
        \dot{v}_1^b = \omega_3^b v_2^b + \sfrac{F_1^b}{m}
    \end{equation}
    \begin{equation}
        \dot{v}_2^b = -\omega_3^bv_1^b + \sfrac{F_2^b}{m}
    \end{equation}
    \begin{equation}
        \dot{\omega}_3^b = \sfrac{K_3^b}{I_3^b}.
    \end{equation}    
\end{subequations}

%%%%%%%%%%%%%%%%%%%%%%%%%%%%%%%%%%%%%

\subsection{Tait-Bryan Angle Surface} \label{sec:Tait_Bryan_surface}
\begin{figure}
    \centering
    \begin{tikzpicture}[inner sep = 0pt, scale = 1]
        \begin{axis}[axis line style={draw=none},
              view={135}{20}, %{35.264},
              tick style={draw=none},
              xticklabels={,,},
              yticklabels={,,},
              zticklabels={,,},
              xmin = -1,
              xmax = 1,
              ymin = -1,
              ymax = 1,
              zmin = -1,
              zmax = 1,
              clip=false
                ]
        
        \draw[thin, ->] (axis cs: 0,0,0) -- (axis cs: 1.5,0,0);
        \draw[thin, ->] (axis cs: 0,0,0) -- (axis cs: 0,1.5,0);
        \draw[thin, ->] (axis cs: 0,0,0) -- (axis cs: 0,0,1.5);
        
        \node[anchor = east] (e1) at (axis cs: 1.5,0,0){\scalebox{0.7}{$\boldsymbol{e}_1^g$}};
        \node[anchor = west] (e2) at (axis cs: 0,1.5,0){\scalebox{0.7}{$\boldsymbol{e}_2^g$}};
        \node[anchor = west] (e3) at (axis cs: 0,0,1.5){\scalebox{0.7}{$\boldsymbol{e}_3^g$}};
        
        \draw[very thin, -] (axis cs: 0,0,0) -- (axis cs: {cos(70)},{sin(70)},0);
        \draw[bend right, ->] (axis cs: .5, 0, 0) to (axis cs: {0.5*cos(70)},{0.5*sin(70)},0) ;
        \node[anchor = west] (a)  at (axis cs: 0.7,0.6,0) {\scalebox{0.7}{$a$}};
        
        \draw[thick, ->] (axis cs: 0,0,0) -- (axis cs: {1.4*cos(70)*cos(35)},{1.4*sin{70}*cos(35)},{1.4*sin(35)});
        \node[anchor = west] (es) at (axis cs: {1.4*cos(70)*cos(35)},{1.4*sin{70}*cos(35)},{1.4*sin(35)}) {\scalebox{0.7}{$\boldsymbol{e}_s^c$}};
        \draw[bend right, ->] (axis cs: {0.6*cos(70)},{0.6*sin(70)},0) to (axis cs: {0.9*cos(70)*cos(35)},{0.9*sin{70}*cos(35)},{0.9*sin(35)}) ;
        \node[anchor = west] (b)  at (axis cs: {1.0*cos(70)*cos(25)},{1.0*sin{70}*cos(25)},{1.0*sin(25)}) {\scalebox{0.7}{$b$}};
        
        \draw[very thin, -] (axis cs:0,0,0) -- (axis cs: {-1.0*sin(70)}, {1.0*cos(70)},0);
        \draw[thick, ->] (axis cs:0,0,0) -- (axis cs: {-1.0*sin(70)*cos(20)}, {1.0*cos(70)*cos(20)},{-1.0*sin(20)});
        \node[anchor = west] (ey) at (axis cs: {-1.0*sin(70)*cos(20)}, {1.0*cos(70)*cos(20)},{-1.0*sin(20)}) {\scalebox{0.7}{$\boldsymbol{e}_y^c$}};
        \draw[bend left, ->] (axis cs: {-0.7*sin(70)}, {0.7*cos(70)},0) to (axis cs: {-0.75*sin(70)*cos(20)}, {0.75*cos(70)*cos(20)},{-0.75*sin(20)}) ;
        \node[anchor = west] (c)  at (axis cs: {-0.8*sin(70)*cos(10)}, {0.8*cos(70)*cos(10)},{-0.8*sin(10)}) {\scalebox{0.7}{$c$}};

        \draw[thick, ->] (axis cs:0,0,0) -- (axis cs: {-0.5},{-0.3},{0.8});
        
        \node[anchor = west] (en)  at (axis cs: {-0.5},{-0.3},{0.8}) {\scalebox{0.7}{$\boldsymbol{e}_n^c$}};
        
        \end{axis}

    \end{tikzpicture}
    \caption{Tait-Bryan Angles}
    \label{fig:tait_bryan}
\end{figure}
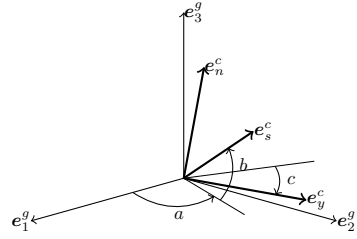

Nonplanar Frenet-like frames can be obtained by adding torsion and normal curvature terms\footnote{The canonical examples are the ``Frenet-Serret" frame~\cite[p.~19]{differential_geometry_of_curves_and_surfaces} and the ``Darboux" frame~\cite[p.~261]{differential_geometry_of_curves_and_surfaces}} to \eqref{eq:frenet_curvature}. However, these frames are difficult to apply to roads since relating curvature to slope and bank is nontrivial. Instead, we propose a nonplanar road parameterization where the {tangent $\boldsymbol{e}_s^c$, normal $\boldsymbol{e}_n^c$ and binormal $\boldsymbol{e}_y^c$ vectors of the centerline} are given by three Tait-Bryan angles $\left\{a,b,c\right\}(s)$ (See Figure \ref{fig:tait_bryan}) and a rotation matrix $\mathbf{R}$ from the global frame to the road surface orientation:
\begin{subequations} \label{eq:tait_bryan_angles}
    \begin{equation}
        \mathbf{R}_a(s) = 
        \begin{bmatrix}
        \cos(a(s)) & -\sin(a(s)) & 0 \\
        \sin(a(s)) & \cos(a(s)) & 0\\
        0 & 0 & 1 
        \end{bmatrix}
    \end{equation}
    \begin{equation}
        \mathbf{R}_b(s) = 
        \begin{bmatrix}
        \cos(b(s)) & 0 & -\sin(b(s)) \\
        0 & 1 & 0 \\
        \sin(b(s)) & 0 & \cos(b(s))
        \end{bmatrix}
    \end{equation}
    \begin{equation}
        \mathbf{R}_c(s) = 
        \begin{bmatrix}
        1 & 0 & 0 \\
        0 & \cos(c(s)) & -\sin(c(s)) \\ 
        0 & \sin(c(s)) & \cos(c(s))
        \end{bmatrix}
    \end{equation}
    \begin{equation}
        \mathbf{R}(s) = 
        \begin{bmatrix}
        \boldsymbol{e}_s^c & \boldsymbol{e}_y^c & \boldsymbol{e}_n^c
        \end{bmatrix}(s)
        =\mathbf{R}_a \mathbf{R}_b  \mathbf{R}_c(s).
    \end{equation}
\end{subequations}
    
As a result we call $a$ the road heading angle, $b$ the road slope angle and $c$ the road bank angle. This is illustrated in Figure \ref{fig:tait_bryan}. As with the Frenet frame, \eqref{eq:general_centerline} and \eqref{eq:general_centerline_ode} define a parametric surface from which we can compute equations of motion. Note that $\boldsymbol{e}_s^c$ and $\boldsymbol{e}_y^c$ are orthonormal so we can use \eqref{eq:simplified_J}. 

%%%%%%%%%%%%%%%%%%%%%%%%%%%%%%%%%%%%%%%%%%%%%%%%%%%%%%%%%%%%
\section{Vehicle Models} \label{sec:Vehicle_Models}

\subsection{Kinematic Vehicle Model}
We use the kinematic bicycle model from~\cite[p.~26]{rajamani_book} with front steering only. Since our equations of motion are about the center of mass, the kinematic constraints are:
\begin{subequations} \label{eq:kinematic_bicycle_constraints}
    \begin{equation}
        \beta = \arctan \left( \frac{l_r}{l_r + l_f}\tan(\gamma)\right)
    \end{equation}
    \begin{equation}
        \omega_3^b = \frac{v\cos(\beta)}{l_r+l_f}\tan(\gamma)
    \end{equation}
    \begin{equation}
        v_1^b = v\cos(\beta)
    \end{equation}
    \begin{equation}
        v_2^b = v\sin(\beta)
    \end{equation}
    \begin{equation}
        \dot{v} = a_t + a_g,
    \end{equation}
\end{subequations}
where $\beta$ is the slip angle of the vehicle. $a_t$ and $\gamma$ are traction control acceleration and steering angle and are treated as inputs to the vehicle, $v$ is the velocity of the vehicle, $l_f$ is the distance from the vehicle COM to the front axle in the direction of $\boldsymbol{e}_1^b$ and $l_r$ is the distance to the rear axle from the COM in the opposite direction. Since $a_t$ is an input to a traction controller and not the total acceleration of the vehicle we add gravitational acceleration\footnote{Aerodynamic drag forces may be considered too; we ignore drag since it differs little between planar and nonplanar environments} $a_g$ which is the component of gravity ($\boldsymbol{g}$) in the direction defined by the slip angle $\beta$: $a_g = \boldsymbol{g} \cdot \left(\boldsymbol{e}_1^b \cos\beta  +\boldsymbol{e}_2^b\sin\beta \right)$. We assume that gravity has magnitude $g$ in the $-\boldsymbol{e}^g_3$ direction. This allows us to express $a_g$ as (see Appendix {\ref{app:gravity_derivation}}):
\begin{equation}\label{eq:gravity_acceleration}
    a_g = - g 
    \begin{bmatrix}
        \dfrac{\boldsymbol{x}^p_s \cdot \boldsymbol{e}_3^g}{\boldsymbol{x}^p_s \cdot \boldsymbol{x}^p_s} &
        \dfrac{\boldsymbol{x}^p_y \cdot \boldsymbol{e}_3^g}{\boldsymbol{x}^p_y \cdot \boldsymbol{x}^p_y}
    \end{bmatrix}
    \mathbf{J}
        \begin{bmatrix}
            \cos(\beta) \\ \sin(\beta)
        \end{bmatrix}.
\end{equation}

Substituting \eqref{eq:kinematic_bicycle_constraints} and \eqref{eq:gravity_acceleration} into the vehicle model from Section \ref{sec:constrained_equations_of_motion} we obtain our kinematic vehicle model:
\begin{subequations} \label{eq:kinematic_vehicle_model}
    \begin{align}
    \begin{split}
        \beta &= \arctan \left( \frac{l_r}{l_r + l_f}\tan(\gamma)\right)
    \end{split}\\
    \begin{split}
        \dot{v} &= a_t - g 
        \begin{bmatrix}
            \dfrac{\boldsymbol{x}^p_s \cdot \boldsymbol{e}_3^g}{\boldsymbol{x}^p_s \cdot \boldsymbol{x}^p_s} &
            \dfrac{\boldsymbol{x}^p_y \cdot \boldsymbol{e}_3^g}{\boldsymbol{x}^p_y \cdot \boldsymbol{x}^p_y}
        \end{bmatrix}
        \mathbf{J}
        \begin{bmatrix}
            \cos(\beta) \\ \sin(\beta)
        \end{bmatrix}
    \end{split}\\
    \begin{split}
        \begin{bmatrix}
        \dot{s}\\
        \dot{y}
        \end{bmatrix}
        &=
        \mathbf{I} ^{-1}
        \mathbf{J}
        \begin{bmatrix}
            \cos(\beta) \\
            \sin(\beta)
        \end{bmatrix} v
    \end{split}\\
    \begin{split}
        \dot{\theta}^s &= \frac{v \cos(\beta)}{l_f + l_r}\tan(\gamma) + 
        \frac{\left(\boldsymbol{x}^p_{ss}\times \boldsymbol{x}^p_s\right)\cdot \boldsymbol{e}_n^p}{\boldsymbol{x}^p_s \cdot \boldsymbol{x}^p_s}\dot{s} 
        \\ &
        + \frac{\left(\boldsymbol{x}^p_{sy}\times \boldsymbol{x}^p_s\right)\cdot \boldsymbol{e}_n^p}{\boldsymbol{x}^p_s \cdot \boldsymbol{x}^p_s}\dot{y}.
    \end{split}
    \end{align}
\end{subequations}
Finally, we can compute the total normal force on the vehicle using \eqref{eq:F3_constraint} and the kinematic constraints. We only consider gravity and a normal force hence
$F_3^b = F_N^b + m(\boldsymbol{g} \cdot \boldsymbol{e}_n^p)$. Using \eqref{eq:shape_operator}, \eqref{eq:F3_constraint} and \eqref{eq:kinematic_bicycle_constraints}:
\begin{equation} \label{eq:normal_force}
    \begin{split}
        F_N^b = m v^2
    \begin{bmatrix}
        \cos\beta & \sin\beta
    \end{bmatrix}
    \mathbf{J}^{-1} \mathbf{II} \ \mathbf{I}^{-1} \mathbf{J}
    \begin{bmatrix}
        \cos\beta \\ \sin\beta
    \end{bmatrix}
    \\
    + (\boldsymbol{e}_n^p \cdot \boldsymbol{e}_3^g) mg.
    \end{split}
\end{equation}
Though not incorporated in kinematic models, maintaining a positive vehicle normal force is necessary for controlling a vehicle. We use this expression as a constraint in Section \ref{sec:mpc}.

\subsection{Dynamic Vehicle Model}
Dynamic models with tire forces can be developed from the framework of section \ref{sec:Motion Along a Surface} by computing wheel slip as a function of vehicle state. However, tire models require modeling of tire normal force which depends on suspension geometry~\cite[Ch.~3]{guiggiani_book} and is not covered by this work. 

\subsection{Model Implementation}
Implementing nonplanar equations of motion is tedious and best-suited for symbolic computation software; we outline our approach below for the kinematic vehicle model and a Tait-Bryan angle surface parameterization:
\begin{enumerate}
    \item Choose $a(s)$, $b(s)$ and $c(s)$ or fit them to data. 
    \item This defines $\boldsymbol{e}^c_s(s)$, $\boldsymbol{e}^c_y(s)$ and $\boldsymbol{e}^c_n(s)$ through \eqref{eq:tait_bryan_angles}
    \item Use \eqref{eq:centerline_partial_derivatives} to compute $\boldsymbol{x}^p_s$ and other partial derivatives. 
    \item {Use {\eqref{eq:normal_vector}} to compute $\boldsymbol{e}^p_n$.}
    \item Use \eqref{eq:parametric_velocity}, \eqref{eq:surface_angular_velocity_unsimplified} and \eqref{eq:simplified_J} to compute $\mathbf{I}$, $\mathbf{II}$ and $\mathbf{J}$
    \item Now use \eqref{eq:kinematic_vehicle_model} to cast in the form $\dot{z} = f(z,u)$, where $z = [v,s,y,\theta^s]$, $u = [a_t,\gamma]$ and $f$ is the nonlinear equation of motion.
\end{enumerate}

\section{Control} \label{sec:control}
\subsection{Nonplanar Model Predictive Control} \label{sec:mpc}
We used the Model Predictive Control (MPC) problem from~\cite{7225830} for following the center of a curved lane at constant speed with minor algorithmic changes:
\begin{itemize}
    \item We replaced the vehicle model used in \cite{7225830} with a fourth order Runge-Kutta integration of the nonplanar kinematic vehicle model \eqref{eq:kinematic_vehicle_model}.
    \item We removed the input magnitude cost so that the controller can track a reference speed while moving uphill. Smooth response was obtained by tuning state and input rate costs. 
    \item We increased the horizon to 20 points which we space 0.05 seconds apart.
\end{itemize}

Additionally, we used an MPC-based speed planner to adjust the target speed of the vehicle to satisfy normal force constraints on upcoming road terrain using \eqref{eq:normal_force}. This was done by solving an optimization problem to find the speed closest to the desired speed for which the vehicle normal force would be in a specified range for the next $30\si{\meter}$ of the road. Due to dependence of normal force on vehicle steering angle and orientation, a rate penalty on adjusted target speed was added to avoid numerical instability of the planner. This planner was run at each time step before solving the nonplanar MPC problem with the adjusted target speed. We used the normal force range $\left[8\si{\kilo\newton}, 40\si{\kilo\newton}\right]$.

\subsection{Comparison Controllers}
We used two other controllers for comparison. First, the same MPC controller but with the planar kinematic Frenet model and no normal force planner. Second, the ``Stanley" controller presented in~\cite{4282788}. We removed the Stanley controller terms for reference yaw rate and steering dynamics since neither is modeled here. The planar MPC and Stanley controllers struggled to make their way up the steeper sections of the simulated road (Figure \ref{fig:road_surface}). To aid this we added a feed-forward term $-\boldsymbol{g}\cdot\boldsymbol{e}^b_1$ to $a_t$ and reapplied input limits before applying to the vehicle.

\section{Simulation}\label{sec:Simulation}
We implemented the kinematic vehicle model \eqref{eq:kinematic_vehicle_model} for the Tait-Bryan surface parameterization from Section \ref{sec:Tait_Bryan_surface} symbolically in CasADi~\cite{Andersson2018}. This automated symbolic computation of $\mathbf{I}$, $\mathbf{II}$, $\mathbf{J}$ and all other terms as functions of vehicle state, input and surface parameterization. We also used CasADi to interpolate and differentiate $a(s)$, $b(s)$ and $c(s)$ in \eqref{eq:tait_bryan_angles} for a user-defined surface, formulate the MPC problem from Section \ref{sec:mpc} and solve the MPC problem using IPOPT~\cite{ipopt}. We used IDAS~\cite{hindmarsh2005sundials} to integrate the kinematic equations of motion \eqref{eq:kinematic_vehicle_model} to simulate the vehicle. Since the MPC problem was run in a receding horizon manner, we warm-started IPOPT with the solution from the previous iteration. For illustrative purposes we did not limit the solve time of IPOPT. We then compared the nonplanar MPC, planar MPC and Stanley controller using the nonplanar kinematic model as a simulator.

We ran all simulations on an Intel Xeon E3-1505M v6 CPU @ 3.00GHz with a simulation time step of 0.05s. We used vehicle model parameters listed in Table \ref{tab:vehicle_parameter_table}.

\setlength{\tabcolsep}{0.5em} % for the horizontal padding
{\renewcommand{\arraystretch}{1.2}% for the vertical padding
\begin{table}[htbp] 
\caption{Vehicle Parameters} \label{tab:vehicle_parameter_table}
\begin{center}
\begin{tabular}{|c|c|}
\hline
\textbf{Parameter} & \textbf{Value} \\
\hline 
$m$  & 2303 \si{\kilo\gram} \\
$l_f$ & 1.52 \si{\metre} \\
$l_r$ & 1.50 \si{\metre} \\
%$h$  & 0.592 \si{\metre} \\
%$I_1^b$ & 956 \si{\kilo\gram\metre\squared} \\
%$I_2^b$ & 5000 \si{\kilo\gram\metre\squared}\\
%$I_3^b$ & 5520 \si{\kilo\gram\metre\squared}\\
\hline
$a_t$ & $\in[-10,10] \si{\metre\per\second\squared}$\\
$\gamma$ & $\in [-.5,.5] \si{\radian}$\\
\hline
\end{tabular}
\label{tab1}
\end{center}
\end{table}
}

\begin{figure} 
\centering
\begin{subfigure}{0.45\linewidth}
    \centering
    \includegraphics[width=0.95\linewidth]{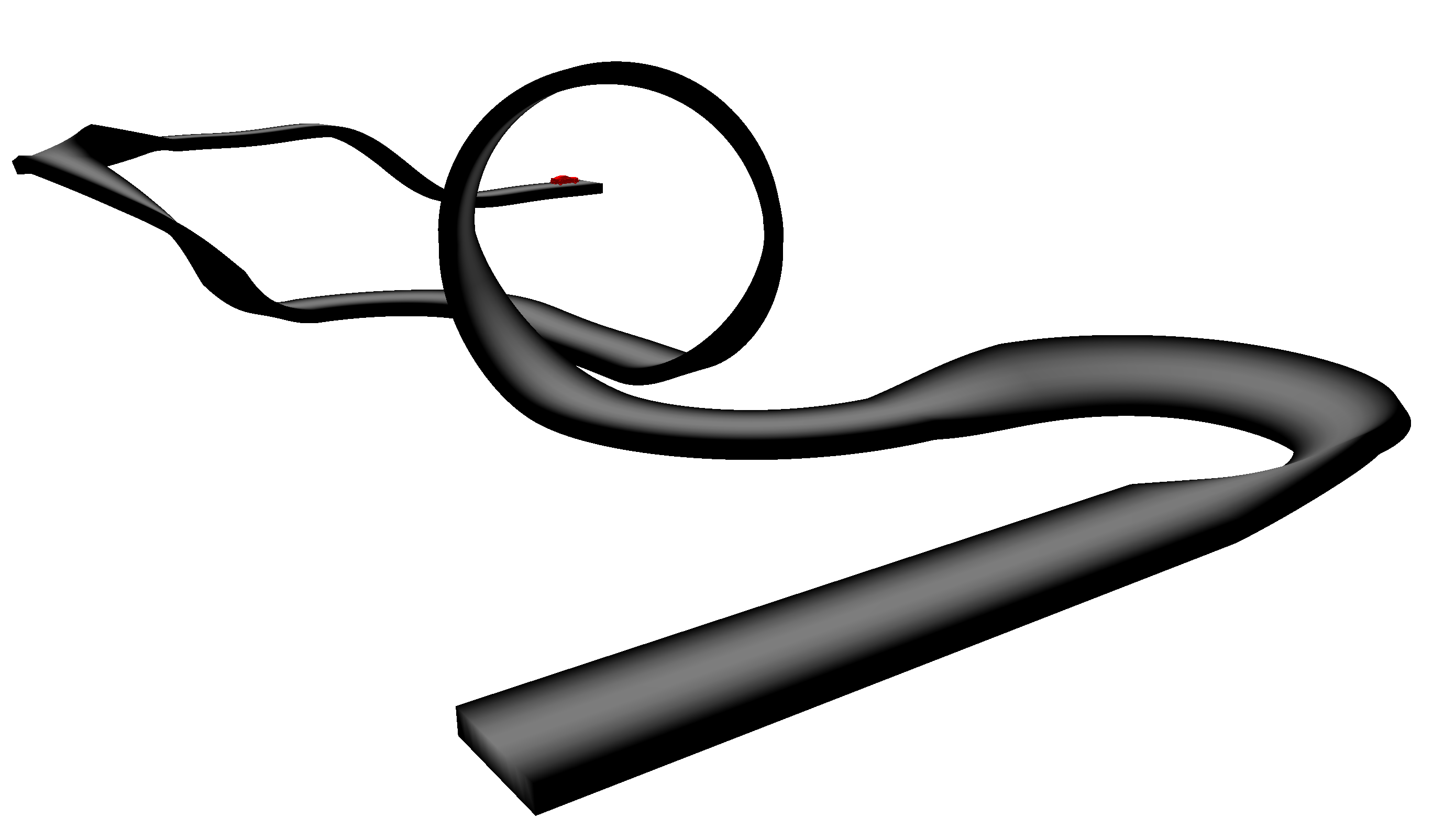}
    \caption{Simulated Road Surface}
    \label{fig:road_surface}
\end{subfigure}
\begin{subfigure}{0.45\linewidth}
    \centering
    \includegraphics[width=0.95\linewidth]{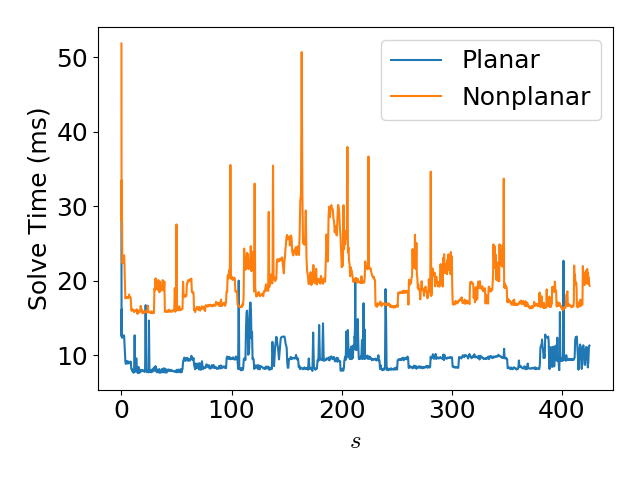}
    \caption{MPC Solve Time}
    \label{fig:solve_time}
\end{subfigure}
\caption{Road Surface and MPC Solve Time. Nonplanar includes solve time for planner. Videos can be found at \href{https://youtu.be/KHP0uUW4oHY}{https://youtu.be/KHP0uUW4oHY}.}
\label{fig:road and solve time}
\end{figure}

The same model was used for both simulation and control. We tested all controllers on the road surface seen in Figure \ref{fig:road_surface} with a vehicle center of mass $0.592\si{\meter}$ above the road. We chose a surface where the centerline was feasible with the stated vehicle parameters and kinematic model. All controllers were tasked with maintaining a vehicle speed of $10 \si{\metre\per\second}$ while following the surface centerline ($y = 0)$. We chose a reference speed that was too slow to traverse the loop without losing road contact and too fast for the banked and sloped sections at the end. 

\section{Results}\label{sec:Results}
Planar and nonplanar MPC controller solve time is shown in Figure \ref{fig:solve_time}. We add the solve time of the planner to the nonplanar MPC controller. The average solve time was $19.3\si{\milli\second}$ for the nonplanar MPC controller and $9.4\si{\milli\second}$ for the planar controller. Source code for the simulations can be found at \href{https://github.com/thomasfork/Nonplanar-Vehicle-Control}{https://github.com/thomasfork/Nonplanar-Vehicle-Control}

\begin{figure*}
\centering
\includegraphics[width=.9\linewidth]{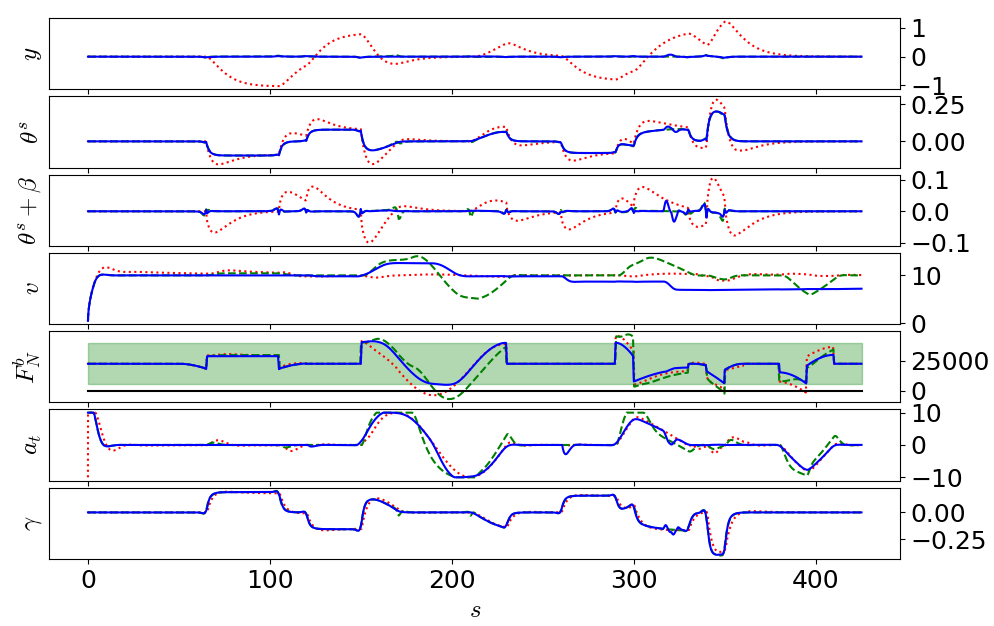}
\caption{Tracking Results for Stanley (dotted red), planar MPC (dashed green) and nonplanar MPC (blue). Shaded normal force region indicates constraints and the solid black line indicates zero. All units are MKS with angles in radians. }
\label{fig:tracking_results}
\end{figure*}

We plot trajectories of position error, heading error, vehicle speed, input and normal force for all controllers in Figure \ref{fig:tracking_results}. Since our equations of motions are written about the center of mass we also plot an adjusted heading error $\beta + \theta^s$, which is the angle between the centerline and vehicle velocity. Planar and nonplanar MPC controllers had comparable lane-following performance but the planar MPC struggled to follow the target speed and routinely lost contact with the road. Furthermore, the Stanley controller accurately tracked the target speed but struggled to follow the centerline and maintain road contact. Our nonplanar MPC controller reliably tracked the centerline while maintaining road contact.

\section{Conclusion}\label{sec:Conclusion}
We developed a new kinematic vehicle model for driving on complex 3D roads modeled by a parametric surface. In doing so, we generalized planar curvilinear vehicle models used for path planning and control. We then designed a model predictive controller for the proposed vehicle model and compared it to existing controllers based on planar vehicle models. Furthermore, we augmented our controller with a planning stage to maintain vehicle-road contact in extreme environments. We note that the proposed modeling approach may be applied to vehicles with suspensions by modeling each tire as a rigid body in tangent contact with additional suspension and vehicle body dynamics. 
\section{Acknowledgements}
{We thank Professor David Limebeer for his feedback on the original submission of this manuscript and highlighting an error in the normal vector calculation $\boldsymbol{e}^p_n$ for the Tait-Bryan angle surface and the sign of several terms in the derivation of $\dot{\theta}^s$.}
\printbibliography

\newpage
\onecolumn
\appendix

\pdfstringdefDisableCommands{\def\eqref#1{(\ref{#1})}}

\subsection{Expanded Derivation From \eqref{eq:theta_s_nasty} to \eqref{eq:theta_s_unsimplified}} \label{app:theta_s}

Starting with \eqref{eq:theta_s_nasty}:
\begin{equation} \nonumber
        \dot{\theta}^s = \omega_3^b + \frac{\boldsymbol{e}_1^b \cdot \boldsymbol{x}^p_{ss} -(\boldsymbol{e}_1^b\cdot \boldsymbol{x}^p_s)(\boldsymbol{x}^p_s\cdot \boldsymbol{x}^p_{ss})(\boldsymbol{x}^p_s\cdot \boldsymbol{x}^p_s)^{-1}}{\boldsymbol{e}_2^b \cdot \boldsymbol{x}^p_s}\dot{s}
        \\
        +\frac{\boldsymbol{e}_1^b \cdot \boldsymbol{x}^p_{sy} -(\boldsymbol{e}_1^b\cdot \boldsymbol{x}^p_s)(\boldsymbol{x}^p_s\cdot \boldsymbol{x}^p_{sy})(\boldsymbol{x}^p_s\cdot \boldsymbol{x}^p_s)^{-1}}{\boldsymbol{e}_2^b \cdot \boldsymbol{x}^p_s}\dot{y}.
\end{equation}

We recognize that the $\boldsymbol{e}^b_3$ component of $\boldsymbol{x}^p_s$ is zero and the $\boldsymbol{e}^b_3$ components of $\boldsymbol{x}^p_{ss}$ and $\boldsymbol{x}^p_{sy}$ are removed by the dot product with $\boldsymbol{x}^p_s$. We introduce placeholder variables $(a,b,c,d,q,r)$ used only here and only for the sake of concise notation defined such that:
\begin{subequations}
    \begin{equation}
        \boldsymbol{x}^p_s = a\boldsymbol{e}_1^b + b\boldsymbol{e}_2^b
    \end{equation}
    \begin{equation}
        \boldsymbol{x}^p_{ss} = c\boldsymbol{e}_1^b + d\boldsymbol{e}_2^b + \boldsymbol{e}_3^b (\boldsymbol{x}^p_{ss} \cdot \boldsymbol{e}_3^b)
    \end{equation}
    \begin{equation}
        \boldsymbol{x}^p_{sy} = q\boldsymbol{e}_1^b + r\boldsymbol{e}_2^b + \boldsymbol{e}_3^b (\boldsymbol{x}^p_{sy} \cdot \boldsymbol{e}_3^b).
    \end{equation}
\end{subequations}
With this in mind, \eqref{eq:theta_s_nasty} becomes:
\begin{equation}
    \dot{\theta}^s = \omega_3^b + \frac{c - a(ac+bd)\frac{1}{a^2+b^2}}{b}\dot{s} + \frac{q-a(aq+br)\frac{1}{a^2+b^2}}{b}\dot{y}.
\end{equation}
And we can rearrange over several steps:
\begin{equation}
    \dot{\theta}^s = \omega_3^b + \frac{c(a^2+b^2) - ca^2-abd}{b(a^2+b^2)}\dot{s} + \frac{q(a^2+b^2) - qa^2-abr}{b(a^2+b^2)}\dot{y}.
\end{equation}
\begin{equation}
    \dot{\theta}^s = \omega_3^b + \frac{cb^2-abd}{b(a^2+b^2)}\dot{s} + \frac{qb^2-abr}{b(a^2+b^2)}\dot{y}.
\end{equation}
\begin{equation}
    \dot{\theta}^s = \omega_3^b + \frac{cb-ad}{a^2+b^2}\dot{s} + \frac{qb-ar}{a^2+b^2}\dot{y}.
\end{equation}
\begin{equation} \nonumber
        \dot{\theta}^s = \omega_3^b + \frac{(\boldsymbol{e}_1^b\cdot \boldsymbol{x}^p_{ss})(\boldsymbol{e}_2^b\cdot \boldsymbol{x}^p_s) - (\boldsymbol{e}_2^b \cdot \boldsymbol{x}^p_{ss})(\boldsymbol{e}_1^b\cdot \boldsymbol{x}^p_s)}{\boldsymbol{x}^p_s\cdot \boldsymbol{x}^p_s}\dot{s}
        \\
        +\frac{(\boldsymbol{e}_1^b \cdot \boldsymbol{x}^p_{sy})(\boldsymbol{e}_2^b\cdot \boldsymbol{x}^p_s) - (\boldsymbol{e}_2^b \cdot \boldsymbol{x}^p_{sy})(\boldsymbol{e}_1^b\cdot \boldsymbol{x}^p_s)}{\boldsymbol{x}^p_s\cdot \boldsymbol{x}^p_s}\dot{y}.
\end{equation}
  
 Which brings us to \eqref{eq:theta_s_unsimplified}.

\subsection{Derivation of \eqref{eq:gravity_acceleration}} \label{app:gravity_derivation}
Beginning with
\begin{equation}
    a_g= -g \boldsymbol{e}^g_3 \cdot(\boldsymbol{e}_1^b \cos(\beta) + \boldsymbol{e}_2^b \sin(\beta)),
\end{equation}
We expand the dot product using the fact that $\boldsymbol{e}_1^b$ and $\boldsymbol{e}_2^b$ are coplanar with $\boldsymbol{x}^p_s$ and $\boldsymbol{x}^p_y$:

\begin{equation}
    a_g = -g 
    \begin{bmatrix}
        \frac{(\boldsymbol{e}^g_3\cdot\boldsymbol{x}^p_s)(\boldsymbol{x}^p_s\cdot\boldsymbol{e}_1^b)}{\boldsymbol{x}^p_s\cdot\boldsymbol{x}^p_s} + 
        \frac{(\boldsymbol{e}^g_3\cdot\boldsymbol{x}^p_y)(\boldsymbol{x}^p_y\cdot\boldsymbol{e}_1^b)}{\boldsymbol{x}^p_y\cdot\boldsymbol{x}^p_y}
        &
        \frac{(\boldsymbol{e}^g_3\cdot\boldsymbol{x}^p_s)(\boldsymbol{x}^p_s\cdot\boldsymbol{e}_2^b)}{\boldsymbol{x}^p_s\cdot\boldsymbol{x}^p_s} + 
        \frac{(\boldsymbol{e}^g_3\cdot\boldsymbol{x}^p_y)(\boldsymbol{x}^p_y\cdot\boldsymbol{e}_2^b)}{\boldsymbol{x}^p_y\cdot\boldsymbol{x}^p_y}
    \end{bmatrix}
    \begin{bmatrix}
        \cos(\beta) \\
        \sin(\beta)
    \end{bmatrix}
\end{equation}
We recognize that this can be factored into a matrix product:
\begin{equation}
    a_g = -g \begin{bmatrix}
        \frac{\boldsymbol{e}^g_3\cdot\boldsymbol{x}^p_s}{\boldsymbol{x}^p_s\cdot\boldsymbol{x}^p_s}
        &
        \frac{\boldsymbol{e}^g_3\cdot\boldsymbol{x}^p_y}{\boldsymbol{x}^p_y\cdot\boldsymbol{x}^p_y}
    \end{bmatrix}
    \begin{bmatrix}
        \boldsymbol{x}^p_s\cdot\boldsymbol{e}_1^b
        &
        \boldsymbol{x}^p_s\cdot\boldsymbol{e}_2^b
        \\
        \boldsymbol{x}^p_y\cdot\boldsymbol{e}_1^b
        &
        \boldsymbol{x}^p_y\cdot\boldsymbol{e}_2^b
    \end{bmatrix}
    \begin{bmatrix}
        \cos(\beta) \\
        \sin(\beta)
    \end{bmatrix}
\end{equation}

Which is \eqref{eq:gravity_acceleration}. This form has the advantage that $\mathbf{J}$ captures the effect of vehicle orientation; no new notion of orientation needs to be introduced, simplifying numerical implementation.

\subsection{Link Between Tait-Bryan Angles and Darboux Frame} \label{app:darboux}
The basis vectors $\boldsymbol{e}_s^c$, $\boldsymbol{e}_y^c$ and $\boldsymbol{e}_n^c$ in \eqref{eq:tait_bryan_angles} are orthonormal by definition, thus they obey the form:
\begin{equation}
    \frac{\partial}{\partial s}\
    \begin{bmatrix}
    \boldsymbol{e}_s^c\\
    \boldsymbol{e}_y^c\\
    \boldsymbol{e}_n^c
    \end{bmatrix}
    =
    \begin{bmatrix}
    0 & \kappa^n & -\kappa^y \\
    -\kappa^n & 0 & \kappa^s \\
    \kappa^y & -\kappa^s & 0
    \end{bmatrix}
    \begin{bmatrix}
    \boldsymbol{e}_s^c\\
    \boldsymbol{e}_y^c\\
    \boldsymbol{e}_n^c
    \end{bmatrix}
\end{equation}
where $\kappa^s$ is the torsion, $\kappa^y$ is the normal curvature, and $\kappa^n$ is the geodesic curvature of the centerline. This is the Darboux Frame. 

By taking derivatives of $\mathbf{R}(s)$ in \eqref{eq:tait_bryan_angles} with respect to s and expressing the result in the basis $\boldsymbol{e}_s^c$ $\boldsymbol{e}_y^c$ $\boldsymbol{e}_n^c$, we can relate the Tait-Bryan angles to the Darboux Frame:
\begin{subequations}
    \begin{equation}
        \kappa^s = -\frac{\partial}{\partial s}(a)\sin(b) - \frac{\partial}{\partial s}(c)
    \end{equation}
    \begin{equation}
        \kappa^y = -\frac{\partial}{\partial s}(a) \sin(c)\cos(b) 
        +\frac{\partial}{\partial s}(b) \cos(c)
    \end{equation}
    \begin{equation}
        \kappa^n = -\frac{\partial}{\partial s}(a)\cos(b)\cos(c) - \frac{\partial}{\partial s}(b)\sin(c).
    \end{equation}
\end{subequations}
As a result we can derive equations of motion for either frame without loss of generality. 

\subsection{Darboux Frame Surface Equations of Motion} \label{app:darboux_eom}
Using the Darboux Frame for the centerline of a centerline surface, we compute the partial derivatives of the parametric surface:
\begin{subequations}
    \begin{equation}
        \boldsymbol{x}^p_s = (1 - y\kappa^n) \boldsymbol{e}_s^c + y \kappa^s \boldsymbol{e}_n^c
    \end{equation}
    \begin{equation}
        \boldsymbol{x}^p_y = \boldsymbol{e}_y^c
    \end{equation}
    \begin{equation}
        \begin{split}
            \boldsymbol{x}^p_{ss} &=  \boldsymbol{e}_s^c\left[y\left(\frac{\partial}{\partial s} \kappa^n +\kappa^s\kappa^y)\right)\right]  \\
            & +\boldsymbol{e}_y^c\left[\kappa^n + y(-(\kappa^s)^2 - (\kappa^n)^2)\right] \\
            & +\boldsymbol{e}^c_n\left[-\kappa^y + y\left(\frac{\partial}{\partial s}\kappa^s + \kappa^y\kappa^n\right)\right]
        \end{split}
    \end{equation}
    \begin{equation}
        \boldsymbol{x}^p_{ys} = \kappa^s \boldsymbol{e}_n^p - \kappa^n \boldsymbol{e}_s
    \end{equation}
    \begin{equation}
        \boldsymbol{x}^p_{yy} = 0
    \end{equation}
\end{subequations}
By definition: 
\begin{equation}
    \boldsymbol{e}^p_n = \frac{\boldsymbol{x}^p_s \times \boldsymbol{x}^p_y }{||\boldsymbol{x}^p_s \times \boldsymbol{x}^p_y ||} 
\end{equation}
Which we can evaluate:
\begin{equation}
    \boldsymbol{e}^p_n = \frac{-\kappa^s y \boldsymbol{e}_s^c + (1-\kappa^n y)\boldsymbol{e}_n^c}{\sqrt{(\kappa^s)^2y^2 + (1-\kappa^n y)^2}}
\end{equation}
It follows that:
\begin{equation}
    \mathbf{I} =
    \begin{bmatrix}
    (\kappa^s)^2y^2 + (1-\kappa^ny)^2 & 0\\
    0 & 1
    \end{bmatrix}
\end{equation}
\begin{equation}
\begin{split}
     \mathbf{II} = 
    \frac{1}{\sqrt{(\kappa^s)^2y^2 + (1-\kappa^ny)^2}}
    \begin{bmatrix}
    \kappa^s y^2 (\frac{\partial}{\partial s} \kappa^n -\kappa^s\kappa^y) + (-\kappa^y + y(\frac{\partial}{\partial s} \kappa^s + \kappa^y\kappa^n))(1-y\kappa^n) & \kappa^s\\
    \kappa^s & 0
    \end{bmatrix}
\end{split}
\end{equation}
\begin{equation}
    \mathbf{J} = 
    \begin{bmatrix}
    \sqrt{(\kappa^s)^2 y^2 + (1-\kappa^ny)^2}\cos\theta^s & -\sqrt{(\kappa^s)^2 y^2 + (1-\kappa^ny)^2}\sin\theta^s\\
    \sin\theta^s & \cos\theta^s
    \end{bmatrix}
\end{equation}
Computing $\dot{s}$, $\dot{y}$ and $\dot{\theta}^s$:
\begin{subequations}
    \begin{equation}
        \dot{s} = \frac{v_1^b\cos\theta^s - v_2^b\sin\theta^s}{\sqrt{(\kappa^s)^2y^2+(1-\kappa^n y)^2}}
    \end{equation}
    \begin{equation}
        \dot{y} = v_1^b\sin\theta^s + v^b_2\cos\theta^s
    \end{equation}
    \begin{equation}
        \dot{\theta}^s = \omega^b_3 - \frac{(\kappa^n +y(-(\kappa^s)^2 - (\kappa^n)^2))}{\sqrt{(\kappa^s)^2 y^2 + (1-\kappa^ny)^2}}\dot{s}
    \end{equation}
\end{subequations}
Which is \eqref{eq:constrained_eom_parametric_kinematic} simplified for a Darboux Frame Surface. 

\end{document}